\documentclass[%
 reprint,prx,
superscriptaddress,
 amsmath,amssymb,
 aps,
floatfix,
]{revtex4-2}
\usepackage{soul,xcolor}
\usepackage{siunitx}
\usepackage{hyperref}
\usepackage{graphicx}
\usepackage{dcolumn}
\usepackage{bm}
\usepackage{amsmath}
\usepackage{times,mathptmx,amssymb}
\usepackage{xcolor}
\usepackage{siunitx}
\usepackage{soul}
\usepackage{physics}

\newcommand{\VEC}[1]{\mathbf{#1}}

\begin{document}
\setstcolor{red}
\raggedbottom
\title{Excitable dynamics driven by mechanical feedback in biological tissues}
\author{Fernanda Pérez-Verdugo}
\thanks{These authors contributed equally}
\affiliation{Department of Physics, Carnegie Mellon University, Pittsburgh, PA, USA}
\author{Samuel Banks}
\thanks{These authors contributed equally}
\affiliation{Department of Physics, Carnegie Mellon University, Pittsburgh, PA, USA}
\affiliation{Department of Physics, Yale University, New Haven, CT, USA}
\author{Shiladitya Banerjee}
\email{Correspondence: shiladtb@andrew.cmu.edu}
\affiliation{Department of Physics, Carnegie Mellon University, Pittsburgh, PA, USA}

\begin{abstract}
Pulsatory activity patterns, driven by mechanochemical feedback, are prevalent in many biological systems. Here we present a theoretical framework to elucidate the mechanical origin and regulation of pulsatile activity patterns within multicellular tissues. We show that a simple mechanical feedback at the level of individual cells -- activation of contractility upon stretch and subsequent inactivation upon turnover of active elements -- is sufficient to explain the emergence of quiescent states, long-range wave propagation, and traveling activity pulse at the tissue-level. We find that the transition between a propagating pulse and a wave is driven by the competition between timescales associated with cellular mechanical response and geometrical disorder in the tissue. This sheds light on the fundamental role of cell packing geometry on tissue excitability and spatial propagation of activity patterns.
\end{abstract}
\maketitle

\section{Introduction} 

Multicellular systems exhibit a wide range of pulsatile and wave-like patterns during collective migration, development, and morphogenesis~\cite{howard2011turing,bailles2022,hakim2017}. The appearance of these patterns can be attributed to various biochemical factors, depending on the specific phenomenon. These include waves of extracellular signal-related kinase (ERK) \cite{hino2020erk, de2021control}, calcium waves \cite{balaji2017calcium}, periodic assembly and disassembly of myosin motors \cite{he2010tissue,martin2009pulsed}, and the periodic release of chemoattractants \cite{palsson1997selection}. Reaction-diffusion models \cite{turing1952,gierer1972,murray2003,hayden2021mathematical} and cellular automaton models \cite{kessler1993pattern,levine1996positive,grace2015regulation} have been widely used to study the mechanisms underlying biochemical pattern formation in multicellular systems. Mechanochemical patterns, on the other hand, have necessitated the development of new classes of models that integrate mechanical forces with chemical reactions~\cite{murray1988,bois2011pattern,banerjee2015,boocock2021theory,staddon2022pulsatile}. For instance, the coupling of mechanical and chemical processes is particularly relevant in understanding the spatial propagation of contraction patterns in \textit{T. Adhaerens} \cite{armon2018ultrafast}, oscillatory morphodynamics in \textit{Drosophila} amnioserosa tissue \cite{lin2017activation}, collective migration patterns~\cite{boocock2021theory} and mechanical waves in expanding MDCK cell monolayers \cite{serra2012mechanical,banerjee2015,banerjee2019}. However, the role of cellular mechanics and geometry in the propagation of mechanochemical signals remains poorly understood. 



One commonly observed mechanical feedback motif in cells is {\it stretch-induced contraction}, wherein a local stretching deformation triggers the recruitment of active components that induce contraction~\cite{odell1981,hino2020erk,heer2017tension,fernandez2009,banerjee2020}. Recent studies have utilized the concept of stretch-induced contraction to elucidate phenomena such as wave propagation in active elastic media~\cite{banerjee2015,banerjee2019}, contraction pulses in epithelial tissues~\cite{armon2021modeling}, cell migration patterns in vitro~\cite{boocock2021theory,boocock2023interplay}, cell and tissue morphogenesis~\cite{lin2017activation,noll2017}. Specifically, all these studies focused on dynamics in active elastic media, without considering the effects of geometric disorder and viscous dissipation on mechanochemical signal propagation. 


In this study, we ask how cellular viscoelasticity and packing geometry regulate the propagation of active stresses at multicellular scales. To this end, we extended the framework of the cellular vertex models~\cite{nagai2001dynamic,farhadifar2007influence,staple2010mechanics,fletcher2014} to incorporate feedback between cell junction strain and contractility. In addition, viscous dissipation is implemented by continuous strain relaxation in cell junctions. We implement a simple feedback rule in which contractility in cell junctions is activated above a threshold junctional stretch. The junction remains active for a duration commensurate with the turnover rate of active elements. This is followed by a refractory period during which junction contractility remains inactive due to the presence of inhibitors of contractility. 
As a result of these rules, each cell junction behaves as an excitable unit that can exist in one of three states: active, inactive, and refractory. 

Our proposed model elucidates the emergence of long-range propagation of contractile pulses and different patterns of self-sustained traveling waves, such as circular, elliptic, and spiral waves. We show that these tissue-level propagation patterns are controlled by the competition between the timescales associated with active and refractory states of the junction, and the characteristic timescale of junction strain relaxation. To explain these observations analytically, we develop an effective theory of coupled excitable junctions, capable of explaining the emergence of the quiescent, wave-like and pulse-like patterns observed in vertex model simulations. Our theoretical framework predicts that shorter junctions promote re-activation of contractility, while larger junctions facilitate the propagation of activity over a broader region of the parameter space. We validate these predictions through simulations of disordered tissues in two dimensions. We find that geometrical disorder promotes sustained wave propagation at the tissue-level, and that the ability of junctions to locally propagate activity increases with its length.


\section{Vertex Model with mechanical feedback}

\subsection{Equations of motion}
To elucidate the emergent dynamic patterns in an excitable tissue, we use the framework of the vertex model~\cite{nagai2001dynamic,farhadifar2007influence,staple2010mechanics,fletcher2014}, where a monolayer tissue is modeled as a two-dimensional polygonal tiling. The polygons represent the cells, and the edges represent the cell-cell junctions. Each vertex $i$, with position $\VEC{r}_i$, is subject to friction with coefficient $\mu$, and elastic forces and inter-cellular tensions arising from a Hamiltonian $H$. The Hamiltonian governing tissue mechanical energy is given by
\begin{equation}
    H=\frac{K}{2}\sum_{\alpha}{(A_\alpha-A_0)^2}
     +\sum_{ \langle i,j\rangle}{\Lambda_{}l_{ij}}\;,
     \label{eq:hamiltonian}
\end{equation}
%
where the first energy term is a sum over all cells $\alpha$, and the second term is a sum over the cell-cell junctions defined by the adjacent vertices $i$ and $j$. The first term in Eq.~\eqref{eq:hamiltonian} is the elastic energy that penalizes changes in cell area, where $K$ is the bulk elastic modulus, $A_\alpha$ and $A_0$ are the actual and preferred cell areas, respectively. The second term represents an interfacial energy, with tension $\Lambda$ along each cell junction of length $l_{ij}$.

Active contractile forces arise at each junction from the actomyosin cortex, generating an active force per unit length, $\Gamma_{ij}(t)$. Consequently, the active force at each vertex can be written as: $\VEC{F}_i^{\text{act}}=-\sum_{\langle i,j\rangle}{\Gamma_{ij}(t)l_{ij}} \left(\partial l_{ij} /\partial \VEC{r}_i\right)$.
As opposed to existing vertex models, here we consider a time-dependent contractility $\Gamma_{ij}(t)$, whose dynamics depend on junctional strain and memory of mechanical state. 
To compute the time-evolution of each vertex, we assumed an overdamped limit, such that the equations of motion are given by:
\begin{equation}
    \mu\frac{{\rm d}{\vb{r}_i}}{{\rm d} t}=-\frac{\partial H}{\partial \VEC{r}_i} +  \VEC{F}_i^{\text{act}}\;.
    \label{eq:eom}
\end{equation}
The above equation of motion is coupled to the dynamics of junctional strain and contractility, as described below.

\begin{figure}[t]
    \centering
    \includegraphics[width=\columnwidth]{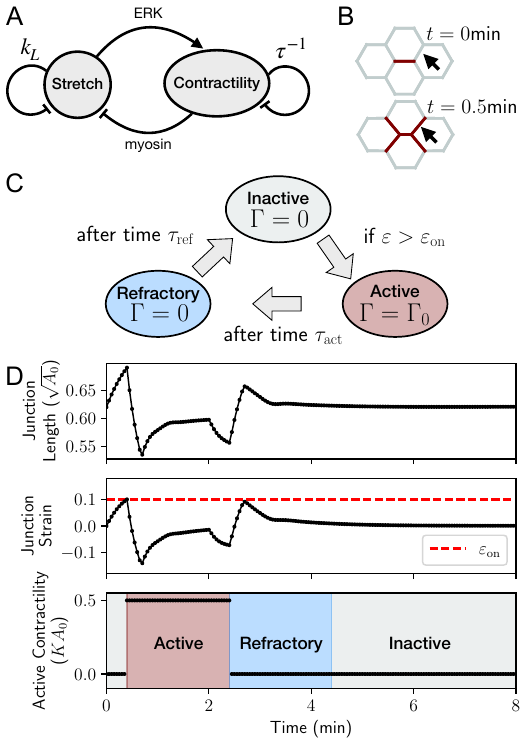}
    \caption{Model for stretch-induced contraction and activity dynamics in cell junctions. (A) Junction stretch induces contractility via ERK activation, while contraction reduces stretch. The mechanical strain relaxes at a rate $k_L$, and the active contraction pulse has a lifetime $\tau$. (B) Representative section of a simulated tissue with hexagonal cells, using $k_L=0.5$ and $\tau=2$. Red junction color denotes active state, which spreads to neighboring junctions as they are stretched. (C) Junction-level rules to change its state. Gray junction color denotes inactive state, while blue denotes a refractory state. (D) Representative dynamics of junction length, strain and active contractile for an initially inactive junction (gray), indicated by the arrow in panel (B), with $\varepsilon_{\text{on}}=0.1$.} 
    \label{fig:model}
\end{figure}

\subsection{Cell junction dynamics under stretch-induced contractility}
Stretch-induced contractility is a commonly observed regulatory mechanism for controlling the level of active contractile stress in cells~\cite{gustafson2022patterned,hino2020erk,nyitrai2004adenosine,heer2017tension,fernandez2009,odell1981}. A local stretch in cell junctions could trigger actin fiber alignment~\cite{uyeda2011,burla2019,banerjee2020}, myosin recruitment~\cite{fernandez2009} and also the activation of the ERK signaling~\cite{hino2020erk} that would promote contractility. We therefore implement a simple model of cellular junctions as viscoelastic materials subject to a strain-tension feedback (Fig.~\ref{fig:model}A). Here, a local stretch triggers the activation of contractility, which in turn reduces stretch via contractile forces. Additionally, junction strain continuously relaxes over time due to viscous dissipation and contractility undergoes turnover as part of a self-regulatory mechanism.
\begin{figure*}
    \centering
\includegraphics[width=\linewidth]{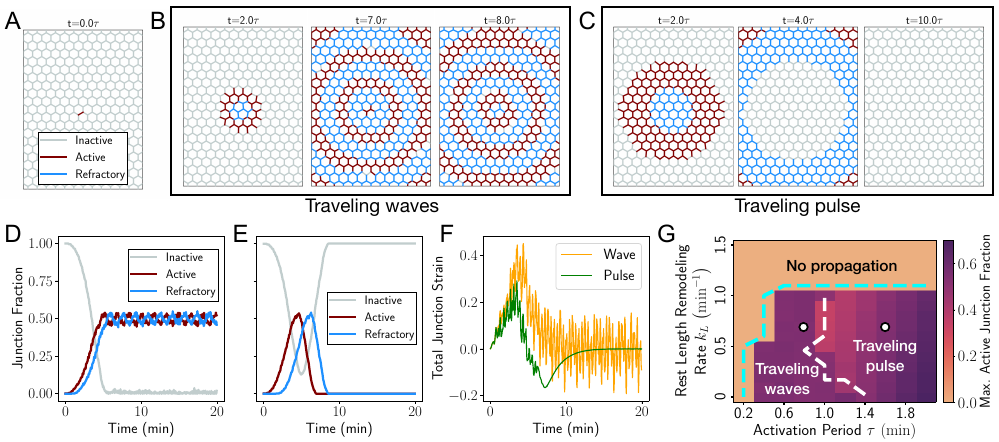}
    \caption{Emergent activity patterns in ordered tissues with a local junction activation. (A) Initial condition of every simulation, with a single chosen junction manually activated (red).
    (B) Snapshots of persistent activity waves in an ordered tissue, in a simulation using the values of $(\tau, k_L)=(0.8,0.7)$ (left white dot in panel (G)). (C) Snapshots of an activity pulse traveling across an ordered tissue, in a simulation using the values of $(\tau, k_L)=(1.6,0.7)$ (right white dot in panel (G)). (D) Fraction of junctions in each state, as a function of time, for activity waves shown in (B). Data show that at long times only active and refractory states persist, with almost no junctions in the inactive state. (E) Fraction of junctions in each state, as a function of time, for activity pulse shown in (B). Data show a transient pulse of activity before the entire tissue becomes inactive. (F) Total junction strain as a function of time, for the wave (B) and the pulse (B). (G) Phase diagram showing the three distinct emergent states: traveling pulse, traveling waves, and a quiescent state with no propagation of activity. The emergent states are controlled by the rest length remodeling rate $k_L$, and the activation period $\tau$.}
    \label{fig:ordered}
\end{figure*}

The mechanical strain in cell junctions is defined as $\varepsilon_{ij}=\left(l_{ij}-l_{ij}^0\right)/{l_{ij}^0}$, where $l_{ij}^0$ is the rest length of the junction shared by the vertices $i$ and $j$. The viscoelastic nature of the junctions is modelled through rest length remodeling at a rate $k_L$, leading to continuous strain relaxation~\cite{staddon2019mechanosensitive}, 
\begin{equation}
\frac{{\rm d}  l_{ij}^0}{{\rm d} t}= -k_L (l_{ij}^0-l_{ij}).
\label{eq:relax}
\end{equation}
Rest length remodeling is a natural consequence of actomyosin networks with turnover, where strained elements are replaced by unstrained ones~\cite{odell1981}. The feedback between junction strain and contractility is implemented as follows. 
Each cell junction can exist in one of three states: {\it inactive} ($\Gamma_{ij}=0$), {\it active} ($\Gamma_{ij}=\Gamma_0$), and {\it refractory} ($\Gamma_{ij}=0$). While both inactive and refractory states lack contractility, refractory junctions are those that cannot be active for a duration $\tau_{\text{ref}}$, representing the timescale associated with the presence of inhibitors of contractility. The rules describing junction state changes are given below (Fig.~\ref{fig:model}C):
\begin{itemize}
    \item {\it Inactive} junctions become active if their strain $\varepsilon_{ij}$ exceeds a threshold value $\varepsilon_{\text{on}}$.
    \item {\it Active} junctions become refractory after being active for a time period $\tau_{\text{act}}$.
    \item {\it Refractory} junctions become inactive after a duration $\tau_{\text{ref}}$.
\end{itemize}

As an example, we'll examine the scenario depicted in Fig.~\ref{fig:model}B-D, where gray, red, and blue junctions represent the inactive, active, and refractory states, respectively. Initially, the junction marked by the black arrow in Fig.~\ref{fig:model}B is set to an inactive state. Contraction of the neighboring junction induce strain levels exceeding the threshold $\varepsilon_{\text{on}}$, thereby triggering ERK signaling activation \cite{hino2020erk}. This activation, in turn, leads to the production of ERK inhibitors \cite{hayden2021mathematical}. The threshold value $\varepsilon_{\text{on}}$ ensures immunity to small perturbations, allowing junctions to be excitable units. ERK activation induces contractility, changing the junction state from inactive to active, increasing contractility to $\Gamma_{ij}=\Gamma_0$. The active state persists for a time period $\tau_{\text{act}}$ (red phase in Fig.~\ref{fig:model}C), which represents an effective timescale arising from the turnover time of actomyosin, and the inactivation of ERK. The remaining levels of ERK inhibitors keep the junction in a refractory state, in which it can not be re-activated (blue phase in Fig.~\ref{fig:model}C). Finally, after a time period $\tau_{\text{ref}}$, the inhibitors reach low enough levels to take the junction back to the inactive state (final state in Fig.~\ref{fig:model}C). 

Although the activation time period ($\tau_{\text{act}}$) and the refractory time period ($\tau_{\text{ref}}$) are distinct parameters, a recent study~\cite{hayden2021mathematical} has demonstrated that, to adequately describe the observed ERK activity waves, the characteristic timescales for ERK activation and inactivation by inhibitors tend to be similar, typically of the order of a few minutes. For simplicity, we will first focus on the case where $\tau_{\text{act}}=\tau_{\text{ref}}=\tau$, and we will refer to this timescale simply as {\it activation period} unless otherwise specified. The strain relaxation rate ($k_L$) and the activation period ($\tau$) will jointly impact the dynamics at both the junction and tissue scales, as elaborated in Sections IIIA-B. Later in Section IIID, we will delve into the effects of varying duration for refractory and active states.





\section{Results} 
\subsection{Traveling pulse and waves in ordered tissues}
To characterize the emergent dynamic states arising from junction-level mechanical feedback, we first simulated an ordered tissue, composed of 260 hexagonal cells, in a box of sides $L_x\sim 14\sqrt{A_0}$ and $L_y\sim 18.6\sqrt{A_0}$, under periodic boundary conditions. In simulations, we non-dimensionalized force scales by $K (A_0)^{3/2}$, length scales $\sqrt{A_0}$, and timescales by $\mu/(K A_0)$, setting $A_0=1$, $K=1$, and $\mu=\SI{0.636 }{(\min)}$.

We initiate our simulations with a mechanically equilibrated tissue, where all junctions are initially in the inactive state. We then perturb the equilibrium state by manually activating a single junction positioned near the center of the simulation window (Fig.~\ref{fig:ordered}A). When the rate of strain relaxation is sufficiently slow, corresponding to a small value of $k_L$, we observe the emergence of two distinct activity patterns depending on the activation period $\tau$. For small $\tau$, we find waves of activity traveling radially outwards, as shown in Fig.~\ref{fig:ordered}B (Movie 1). These self-sustaining waves are characterized by alternating rings and regions of red (indicating activity) and blue (indicating refractory) junctions. The tissue activity reaches a steady-state when the wavefront traverses the entire tissue (around $t\sim 5\tau$ in Fig.~\ref{fig:ordered}D). Conversely, for larger values of $\tau$, we do not observe self-sustaining waves due to the lack of junction reactivation events. Instead, a single transient activity pulse travels across the tissue (Figs.~\ref{fig:ordered}B and D, Movie 2). Over an extended time period, the tissue eventually becomes entirely inactive.

To quantify the mechanical deformation due to these traveling activity patterns, we calculated the total tissue strain as $\varepsilon_{\text{total}} = \sum_{\langle i,j\rangle} \varepsilon_{ij}$. Fig.~\ref{fig:ordered}F shows the dynamics of the total strain for both wave (Fig.~\ref{fig:ordered}A) and pulse-like (Fig.~\ref{fig:ordered}B) patterns. The pulse causes a positive peak in strain, followed by a negative peak, ultimately returning to zero strain due to mechanical relaxation. Conversely, in the traveling wave pattern, while there is a peak in strain, it eventually stabilizes as a result of activity-induced mechanical fluctuations and the relaxation of strain at the junction level.

These propagating activity states are only observed when the value of $k_L$ is sufficiently small. A large $k_L$ causes the strain in the neighboring junctions of an active junction to relax before activation can occur, resulting in a quiescent state without any propagation. To quantify the extent of tissue-scale activity, we calculated the maximum fraction of active junctions throughout the simulation. This measurement enables us to identify the phase boundary, determined by the critical value of $k_L$, that separates the regimes with activity propagation (either wave or pulse) from those without propagation (cyan-dashed boundary in Fig.\ref{fig:ordered}G). Moreover, by quantifying the active junction fraction at the final steady state, we can differentiate between the propagating modes, leading to the delineation of the wave-to-pulse phase boundary (white-dashed boundary in Fig.\ref{fig:ordered}G).

\subsection{Effective theory predicts emergent dynamic states}
\label{subsec:model}
\begin{figure}[t]
    \centering
\includegraphics[width=\columnwidth]{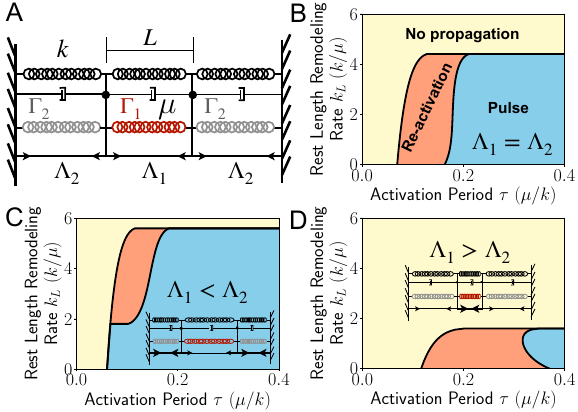}
    \caption{Effective model of coupled excitable junctions. (A) Schematic of a minimal three-junction model. (B) Phase diagram for pulse propagation and reactivation when the junctions have equal tension. (C-D) Phase diagrams for pulse propagation and reactivation for asymmetric junction tensions, with $(\Lambda_1,\Lambda_2=0.1,0.2)$ (C), and $(\Lambda_1,\Lambda_2=0.4,0.1)$ (D).}
    \label{fig:theory}
\end{figure}

To analytically predict the emergence of excitable pulses, quiescent states, and oscillatory patterns as functions of the strain relaxation rate $k_L$ and activation period $\tau$, we developed an effective one-dimensional theory of coupled excitable junctions. Our minimal model consists of three interconnected junctions with fixed boundaries, as shown in Fig. \ref{fig:theory}A. Each unit comprises an elastic component with a spring constant $k$ and natural length $L$ (representing the one-dimensional version of cell elasticity), connected in parallel with a dashpot of friction coefficient $\mu$, and an active element with contractility $\Gamma_{1,2}$. If the junction is inactive or refractory then $\Gamma_{1,2}=0$, and $\Gamma_{1,2}=\Gamma_0$ if the junction is active. These active and elastic elements are connected in parallel with a tensile element with line tension $\Lambda_{1,2}$. The central junction has a length $l_1(t)$, while the outer junctions have lengths  $l_2(t)$. The fixed boundary conditions ensure that $l_1(t)+2l_2(t)=3L$. 

The system is initialized in a mechanical equilibrium state, and we perturb it by activating the central junction ($\Gamma_{1}=\Gamma_0$, $\Gamma_{2}=0$). We then let the system to evolve following the equations of motion: $\mu{\rm d}{l_i}/{{\rm d} t}=-\partial H_{\text{eff}}/\partial l_i$, Eq.~\eqref{eq:relax}, and the rules governing the junction states. The effective Hamiltonian governing the system is defined as:
\begin{align}
    H_{\text{eff}}=&\frac{k}{2}(l_1-L)^2+k(l_2-L)^2+\Lambda_1 l_1 +2\Lambda_2 l_2+ \frac{\Gamma_1}{2}l_1^2+\Gamma_2 l_2^2.
    \label{eq:enegy1D}
\end{align}

We initially considered the scenario of symmetric junctions, wherein $\Lambda_1 = \Lambda_2$. This corresponds to an ordered tissue where junction tensions and lengths are uniform. To explore the behavior of the system, we numerically solve the system of equations for different values of $\tau$ and $k_L$, from $t=0$ to $t=2\tau$. The simulation outcomes can be categorized as follows: i) If the outer junctions remain inactive throughout the simulation, it is classified as a case of {\it No propagation}; ii) If the outer junctions become active but the central junction does not re-activate, we observe a single {\it Pulse}; and iii) finally, if the outer junctions become active and the central junction re-activates, it falls into the category of {\it Re-activation}.
Fig. \ref{fig:theory}B shows the phase diagram of the model in $\tau$-$k_L$ phase space showing the emergence of the three outcomes described above. A comparison with the phase diagram for the ordered tissue (Fig. \ref{fig:ordered}E) reveals that the effective model successfully captures both key features of the vertex model: a critical value of $k_L$ for propagation of activity, which diminishes for small $\tau$, and a small region of reactivation corresponding to wave-like states.

We then used the one-dimensional effective model (Fig.~\ref{fig:theory}A) to investigate the role of disorder in the propagation of activity. Disorder was introduced by removing the condition of homogeneous line tension, letting $\Lambda_1 \neq \Lambda_2$. First, we analyzed the case $\Lambda_1 < \Lambda_2$. Due to identical mechanical properties of each junction before activation (other than tension values), the initial equilibrium state featured a central long junction ($l_1 > L$) flanked by two shorter junctions ($l_2 < L$) (Fig.~\ref{fig:theory}C). By solving the system of equations numerically, we found that the larger junction ($l_1 > L$) could propagate activity over a broader region in the $(\tau, k_L)$ parameter space, while the re-activation region is substantially diminished. 
This is because larger junctions produce greater active contractile forces, while shorter neighboring junctions require a lower extension to achieve the strain threshold for activation $\varepsilon_{\text{on}}$.
Conversely, when $\Lambda_1 > \Lambda_2$, the opposite behavior was observed. Our effective model thus reveals two main effects of the geometrical heterogeneity (or disorder) on cellular response to active contractility. Large junctions promote propagation of activity, while shorter junctions facilitated re-activation, leading to oscillatory patterns.
\begin{figure}[t]
    \centering
\includegraphics[width=\columnwidth]{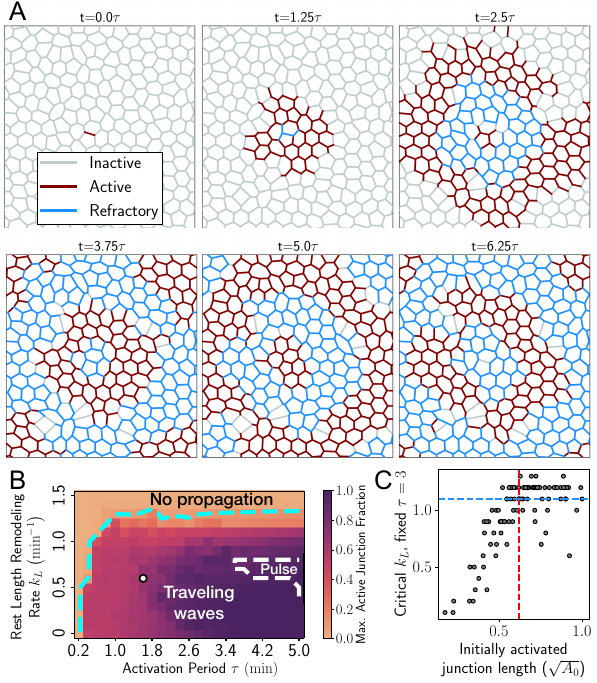}
    \caption{Sustained propagation of activity waves in disordered tissues. (A) Snapshots of activity waves traveling across a disordered tissue, using $(\tau, k_L)=(1.6,0.7)$ (white dot in (B)). Colored segments represent inactive (gray), active (red), and refractory junctions. (B) Phase diagram for the emergence of waves, pulses and quiescent states, varying strain relaxation rate $k_L$ and activation period $\tau$. (C) Scatter plot of the critical rest length remodeling rate $k_L$ for a fixed $\tau=3$, as a function of the initially activated junction length, for 100 different chosen junctions. The critical $k_L$ is defined as the maximum $k_L$ that allow propagating states. 
    Red-vertical line represents the junction length in the ordered hexagonal tissue. Blue-horizontal line represents the critical $k_L$ in the ordered tissue.}
    \label{fig:disordered}
\end{figure}
\subsection{Tissue disorder promotes self-sustained wave propagation}
Motivated by the predictions of the effective model on the impact of geometric heterogeneity, we now investigate the effect of disorder in cell packing geometry on activity propagation in two-dimensional tissue simulations. 
To this end, we constructed a tissue comprising 208 cells, within a rectangular box with dimensions approximately equal to $L_x\sim 14 \sqrt{A_0}$ and $L_y\sim 15\sqrt{A_0}$, subject to periodic boundary conditions. In these simulations, all mechanical properties at cell and junction levels are the same, with disorder restricted to geometric heterogeneity only. The initial state of the tissue corresponded to a state of mechanical equilibrium, characterized by varying junction lengths and polygon sidedness, as depicted in Fig.~\ref{fig:disordered}A.

As previously, we activated a randomly chosen cell junction (see Fig.~\ref{fig:disordered}A, $t=0.0\tau$), and let the tissue evolve from $t=0$ to $t=20$, for different values of strain relaxation rate $k_L \in (0,1.5)$ and activation period $\tau \in (0.2, 5.0)$.
By measuring the maximum active junction fraction (Fig.~\ref{fig:disordered}B), we again observe that propagation occurs below a critical $k_L$, for sufficiently large $\tau$. Unlike in ordered tissues (Fig.~\ref{fig:ordered}G), wave states are now possible for a wide range of $\tau$ values, and propagating solitary pulse only occurs in particular cases with exceedingly large activation periods. 
Consistent with the predictions of the one-dimensional effective model, we find that the presence of short junctions in disordered tissues promotes junction reactivation, thereby facilitating the emergence of self-sustaining wave-like states. As an illustrative example, Fig.\ref{fig:disordered}A (corresponding to the white dot in Fig.\ref{fig:disordered}B) represents a wave-like state arising in a tissue with parameters $(\tau=1.6,k_L=0.7)$ (Movie 3), which led to pulse propagation in the ordered tissue (Fig.\ref{fig:ordered}E). Interestingly, the junction that is reactivated by the end of an oscillatory cycle need not necessarily be the same one initially chosen for activation. This introduces a non-local effect of disorder in promoting sustained wave-like patterns. We find that the critical $k_L$ required for wave propagation increases with the length of the initially activated junction (Fig.\ref{fig:disordered}C), as predicted by the one-dimensional effective model.

\begin{figure}[t]
    \centering
\includegraphics[width=\columnwidth]{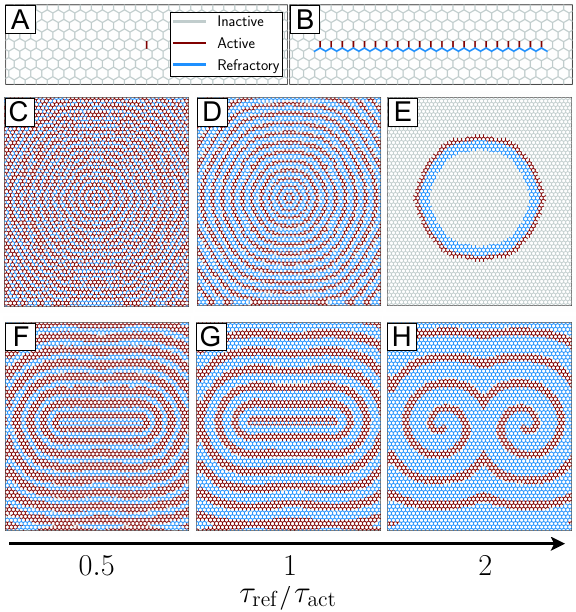}
    \caption{Emergent dynamical patterns arising in an ordered tissue composed of 2759 cells, varying $\tau_{\text{ref}}/\tau_{\text{act}}$. Colored segments represent inactive (gray), active (red), and refractory junctions. (A-B) Zoomed in configuration of two different initial conditions for junction states: (A) single junction activation, and (B) partial row activation with neighbours underneath in refractory states. (C-D-E) Collective dynamics arising from the initial condition in (A), for different values of $\tau_{\text{ref}}$, with $\tau_{\text{act}}=0.6$. (F-G-H) Collective dynamical states arising from the initial condition (B), for different values of $\tau_{\text{ref}}$, with $\tau_{\text{act}}=1$. (C-D-F-G-H) represent self-sustained waves, while (E) represents a solitary traveling pulse. All these simulations consider $k_L=0.5$. }
    \label{fig:patterns}
\end{figure}
\subsection{Controlling the geometry of wavefronts}

Our theory and simulations have elucidated that the propagation of activity at the tissue scale is governed by two distinct characteristic timescales of the system: the activation period $(\tau)$ (taken to be equal to the refractory time) and the rest length remodeling timescale $(k_L^{-1})$. We now investigate the impact of varying the activation ($\tau_{\text{act}}$) and refractory periods ($\tau_{\text{ref}}$) on the resulting dynamic patterns that emerge within the tissue. In particular, we show that the ratio of activation to refractory period controls the geometry of wave patterns.

As in previous simulations, we initiated the simulation by activating a single junction within the tissue, while the remaining junctions remained in the inactive state (see Fig.\ref{fig:patterns}A). We find that the ratio of the activation period to the refractory period, $\Delta = \tau_{\text{act}}/\tau_{\text{ref}}$, controls the wavelength of the propagating waves (see Figs.\ref{fig:patterns}C-D). Specifically, at higher values of $\Delta$, propagating waves fail to materialize, and instead, we observe the presence of a solitary traveling pulse of activity (Fig.~\ref{fig:patterns}E).

Inspired by self-sustained spiral patterns observed in excitable systems~\cite{grace2015regulation,sawai2005,pertsov1993,lechleiter1991}, we inquired whether we could design an initial state that would break the circular symmetry of the emergent wavefronts. Previous theoretical work using a three-state (inactive, active, refractory) cellular automaton model has shown that spiral waves can emerge from an initial state consisting of a layer of excited cells and an adjacent layer of refractory cells~\cite{reshodko1975computer,greenberg1978spatial,gerhardt1990}. We therefore initialized our simulations by activating a partial row of junctions, with the neighbors underneath active junctions initialized in a refractory state (Fig.~\ref{fig:patterns}B). This initial condition leads to elliptical wavefronts for the case of $\Delta=1$ (Fig.~\ref{fig:patterns}G, Movie 4). Similar to the case of single junction activation, smaller values for $\Delta$ decrease the wavelength of the traveling wavefront (Fig.~\ref{fig:patterns}F). For higher values of $\Delta$ we observe the emergence of a pair of self-sustaining spirals (Fig.~\ref{fig:patterns}H, Movie 5). This can be explained as follows. The initial condition of a partial row of active junctions followed by a layer of refractory junctions instigates two distinct pattern formations. Firstly, it leads to the emergence of a propagating wavefront. Secondly, it initiates the formation of two open ends within the wavefront, resulting initially in the addition of an excited element and subsequently in the development of a curved wave segment. This curved segment then propagates outward, adopting a spiral shape. Thus, each open end leads to the emergence of an spiral. Note that due to the periodic boundary conditions in our model, the least number of open ends that can be created is two. These results show that by designing appropriate initial states and disparate timescales for junction activation and refractory periods, the geometry and wavelength of the emergent wavefronts can be precisely controlled in our model.

\section{Conclusions}
In this manuscript, we have introduced a minimal model designed to elucidate the behavior of mechanically-excitable tissues and investigated the role of viscous dissipation and geometrical disorder on tissue-level pattern formation. Traditional approaches for studying dynamic patterns, such as reaction-diffusion equations and cellular automata models, are constrained in their ability to account for spatial deformation in their standard formulations. To address this limitation, we have employed a vertex-based model incorporating agent-based rules at the junction level, representing coarse-grained biochemical reactions that connect junction deformations with the activation of contractility.

Prior work has examined pattern formation in excitable tissues, considering various triggering factors, including cell-level tension~\cite{armon2021modeling}, cell size~\cite{boocock2023interplay}, and active ERK concentration~\cite{hino2020erk}. However, none of these models have considered the effects of viscous dissipation or explored the potential roles of geometrical disorder in pattern formation dynamics. Within our model, which encompasses three distinct junction-level states (active, inactive, and refractory) and considers mechanical strain as the triggering quantity, we observed the emergence of three tissue-level states: quiescent (no propagation), traveling waves, and traveling pulses. These states arise from the interplay between the characteristic timescales associated with junction activation, inactivation, and refractory states.

To explain these emergent dynamics, we have developed an effective junction-scale theory that qualitatively captures the observed behaviors in the vertex model. Our model also provides insights into the impact of geometrical tissue disorder on tissue-level activity states, demonstrating that large junctions promote propagation, while small junctions facilitate re-activation. These predictions have been corroborated through two-dimensional vertex-like simulations, although experimental validation in epithelial tissues remains an avenue for future exploration.


Furthermore, we have demonstrated that the geometry of the emerging traveling wavefronts is influenced by the initial state of junctions and the ratio between the durations of junction active states and refractory states. This intricate interplay results in variations in wavelengths, transitions from waves to pulses, formation of elliptic wavefronts, and pairs of self-sustained spiral wavefronts. The predicted patterns arising from specific initial junction states could potentially be experimentally tested using optogenetic tools to spatially activate myosin contractility~\cite{cavanaugh2020rhoa}, ERK~\cite{toettcher2013}, and FRET-imaging to visualize the resulting patterns~\cite{hino2020erk}.


\begin{table}
\begin{center}
\begin{tabular}{ |c|c|c| } 
\hline
Parameter & Symbol & Value  \\
\hline
Area elastic modulus & $K$ & 1  \\ 
Preferred area & $A_0$ & 1  \\ 
Friction coefficient & $\mu$ & $\SI{0.636 }{(\min)}$ \\ 
Line tension & $\Lambda_0$& 0.1   \\ 
Active contractility & $\Gamma_0$& 0.5   \\ 
Strain relaxation rate & $k_L$& 1   \\ 
Critical strain threshold & $\varepsilon_c$& 0.1   \\ 
Activation strain & $\varepsilon_{\text{on}}$& 0.1   \\ 
Length threshold for T1 event & $l_{T_1}$& 0.05\\ 
Integration time step & $\Delta t$ & 0.01  \\
\hline
\end{tabular}\label{tab.parameters}
\caption{Default model parameters.}
\end{center}
\end{table}

\section*{Methods}
The custom simulation code for the vertex model was implemented using Python~3. Specifically, the simulation code for an ordered tissue featuring traveling wave states can be accessed on GitHub (\href{https://github.com/BanerjeeLab/Excitable_Tissue}{https://github.com/BanerjeeLab}). In implementing T1 transitions, a similar approach to that described in Ref.~\cite{perez2022tension} was adopted. This involves enforcing the creation and instantaneous resolution of a 4-fold vertex whenever a junction's length becomes smaller than $l_{\text{T1}}$. A newly created junction is set to have $l=l^0=1.5l_{\text{T1}}$. The simulations encompassed tissues of varying sizes, as specified in the respective figure captions. Default model parameters used in the simulations are listed in Table~\ref{tab.parameters}. The numerical analysis of the one-dimensional effective model was done in Mathematica 12. The code for this analysis is also available on GitHub (\href{https://github.com/BanerjeeLab/Excitable_Tissue}{https://github.com/BanerjeeLab}).


%


\begin{thebibliography}{50}%
\makeatletter
\providecommand \@ifxundefined [1]{%
 \@ifx{#1\undefined}
}%
\providecommand \@ifnum [1]{%
 \ifnum #1\expandafter \@firstoftwo
 \else \expandafter \@secondoftwo
 \fi
}%
\providecommand \@ifx [1]{%
 \ifx #1\expandafter \@firstoftwo
 \else \expandafter \@secondoftwo
 \fi
}%
\providecommand \natexlab [1]{#1}%
\providecommand \enquote  [1]{``#1''}%
\providecommand \bibnamefont  [1]{#1}%
\providecommand \bibfnamefont [1]{#1}%
\providecommand \citenamefont [1]{#1}%
\providecommand \href@noop [0]{\@secondoftwo}%
\providecommand \href [0]{\begingroup \@sanitize@url \@href}%
\providecommand \@href[1]{\@@startlink{#1}\@@href}%
\providecommand \@@href[1]{\endgroup#1\@@endlink}%
\providecommand \@sanitize@url [0]{\catcode `\\12\catcode `\$12\catcode
  `\&12\catcode `\#12\catcode `\^12\catcode `\_12\catcode `\%12\relax}%
\providecommand \@@startlink[1]{}%
\providecommand \@@endlink[0]{}%
\providecommand \url  [0]{\begingroup\@sanitize@url \@url }%
\providecommand \@url [1]{\endgroup\@href {#1}{\urlprefix }}%
\providecommand \urlprefix  [0]{URL }%
\providecommand \Eprint [0]{\href }%
\providecommand \doibase [0]{https://doi.org/}%
\providecommand \selectlanguage [0]{\@gobble}%
\providecommand \bibinfo  [0]{\@secondoftwo}%
\providecommand \bibfield  [0]{\@secondoftwo}%
\providecommand \translation [1]{[#1]}%
\providecommand \BibitemOpen [0]{}%
\providecommand \bibitemStop [0]{}%
\providecommand \bibitemNoStop [0]{.\EOS\space}%
\providecommand \EOS [0]{\spacefactor3000\relax}%
\providecommand \BibitemShut  [1]{\csname bibitem#1\endcsname}%
\let\auto@bib@innerbib\@empty
\bibitem [{\citenamefont {Howard}\ \emph {et~al.}(2011)\citenamefont {Howard},
  \citenamefont {Grill},\ and\ \citenamefont {Bois}}]{howard2011turing}%
  \BibitemOpen
  \bibfield  {author} {\bibinfo {author} {\bibfnamefont {J.}~\bibnamefont
  {Howard}}, \bibinfo {author} {\bibfnamefont {S.~W.}\ \bibnamefont {Grill}},\
  and\ \bibinfo {author} {\bibfnamefont {J.~S.}\ \bibnamefont {Bois}},\
  }\bibfield  {title} {\bibinfo {title} {Turing's next steps: the
  mechanochemical basis of morphogenesis},\ }\href@noop {} {\bibfield
  {journal} {\bibinfo  {journal} {Nature Reviews Molecular Cell Biology}\
  }\textbf {\bibinfo {volume} {12}},\ \bibinfo {pages} {392} (\bibinfo {year}
  {2011})}\BibitemShut {NoStop}%
\bibitem [{\citenamefont {Bailles}\ \emph {et~al.}(2022)\citenamefont
  {Bailles}, \citenamefont {Gehrels},\ and\ \citenamefont
  {Lecuit}}]{bailles2022}%
  \BibitemOpen
  \bibfield  {author} {\bibinfo {author} {\bibfnamefont {A.}~\bibnamefont
  {Bailles}}, \bibinfo {author} {\bibfnamefont {E.~W.}\ \bibnamefont
  {Gehrels}},\ and\ \bibinfo {author} {\bibfnamefont {T.}~\bibnamefont
  {Lecuit}},\ }\bibfield  {title} {\bibinfo {title} {Mechanochemical principles
  of spatial and temporal patterns in cells and tissues},\ }\href@noop {}
  {\bibfield  {journal} {\bibinfo  {journal} {Annual Review of Cell and
  Developmental Biology}\ }\textbf {\bibinfo {volume} {38}},\ \bibinfo {pages}
  {321} (\bibinfo {year} {2022})}\BibitemShut {NoStop}%
\bibitem [{\citenamefont {Hakim}\ and\ \citenamefont
  {Silberzan}(2017)}]{hakim2017}%
  \BibitemOpen
  \bibfield  {author} {\bibinfo {author} {\bibfnamefont {V.}~\bibnamefont
  {Hakim}}\ and\ \bibinfo {author} {\bibfnamefont {P.}~\bibnamefont
  {Silberzan}},\ }\bibfield  {title} {\bibinfo {title} {Collective cell
  migration: a physics perspective},\ }\href@noop {} {\bibfield  {journal}
  {\bibinfo  {journal} {Reports on Progress in Physics}\ }\textbf {\bibinfo
  {volume} {80}},\ \bibinfo {pages} {076601} (\bibinfo {year}
  {2017})}\BibitemShut {NoStop}%
\bibitem [{\citenamefont {Hino}\ \emph {et~al.}(2020)\citenamefont {Hino},
  \citenamefont {Rossetti}, \citenamefont {Mar{\'\i}n-Llaurad{\'o}},
  \citenamefont {Aoki}, \citenamefont {Trepat}, \citenamefont {Matsuda},\ and\
  \citenamefont {Hirashima}}]{hino2020erk}%
  \BibitemOpen
  \bibfield  {author} {\bibinfo {author} {\bibfnamefont {N.}~\bibnamefont
  {Hino}}, \bibinfo {author} {\bibfnamefont {L.}~\bibnamefont {Rossetti}},
  \bibinfo {author} {\bibfnamefont {A.}~\bibnamefont
  {Mar{\'\i}n-Llaurad{\'o}}}, \bibinfo {author} {\bibfnamefont
  {K.}~\bibnamefont {Aoki}}, \bibinfo {author} {\bibfnamefont {X.}~\bibnamefont
  {Trepat}}, \bibinfo {author} {\bibfnamefont {M.}~\bibnamefont {Matsuda}},\
  and\ \bibinfo {author} {\bibfnamefont {T.}~\bibnamefont {Hirashima}},\
  }\bibfield  {title} {\bibinfo {title} {Erk-mediated mechanochemical waves
  direct collective cell polarization},\ }\href@noop {} {\bibfield  {journal}
  {\bibinfo  {journal} {Developmental cell}\ }\textbf {\bibinfo {volume}
  {53}},\ \bibinfo {pages} {646} (\bibinfo {year} {2020})}\BibitemShut
  {NoStop}%
\bibitem [{\citenamefont {De~Simone}\ \emph {et~al.}(2021)\citenamefont
  {De~Simone}, \citenamefont {Evanitsky}, \citenamefont {Hayden}, \citenamefont
  {Cox}, \citenamefont {Wang}, \citenamefont {Tornini}, \citenamefont {Ou},
  \citenamefont {Chao}, \citenamefont {Poss},\ and\ \citenamefont
  {Di~Talia}}]{de2021control}%
  \BibitemOpen
  \bibfield  {author} {\bibinfo {author} {\bibfnamefont {A.}~\bibnamefont
  {De~Simone}}, \bibinfo {author} {\bibfnamefont {M.~N.}\ \bibnamefont
  {Evanitsky}}, \bibinfo {author} {\bibfnamefont {L.}~\bibnamefont {Hayden}},
  \bibinfo {author} {\bibfnamefont {B.~D.}\ \bibnamefont {Cox}}, \bibinfo
  {author} {\bibfnamefont {J.}~\bibnamefont {Wang}}, \bibinfo {author}
  {\bibfnamefont {V.~A.}\ \bibnamefont {Tornini}}, \bibinfo {author}
  {\bibfnamefont {J.}~\bibnamefont {Ou}}, \bibinfo {author} {\bibfnamefont
  {A.}~\bibnamefont {Chao}}, \bibinfo {author} {\bibfnamefont {K.~D.}\
  \bibnamefont {Poss}},\ and\ \bibinfo {author} {\bibfnamefont
  {S.}~\bibnamefont {Di~Talia}},\ }\bibfield  {title} {\bibinfo {title}
  {Control of osteoblast regeneration by a train of erk activity waves},\
  }\href@noop {} {\bibfield  {journal} {\bibinfo  {journal} {Nature}\ }\textbf
  {\bibinfo {volume} {590}},\ \bibinfo {pages} {129} (\bibinfo {year}
  {2021})}\BibitemShut {NoStop}%
\bibitem [{\citenamefont {Balaji}\ \emph {et~al.}(2017)\citenamefont {Balaji},
  \citenamefont {Bielmeier}, \citenamefont {Harz}, \citenamefont {Bates},
  \citenamefont {Stadler}, \citenamefont {Hildebrand},\ and\ \citenamefont
  {Classen}}]{balaji2017calcium}%
  \BibitemOpen
  \bibfield  {author} {\bibinfo {author} {\bibfnamefont {R.}~\bibnamefont
  {Balaji}}, \bibinfo {author} {\bibfnamefont {C.}~\bibnamefont {Bielmeier}},
  \bibinfo {author} {\bibfnamefont {H.}~\bibnamefont {Harz}}, \bibinfo {author}
  {\bibfnamefont {J.}~\bibnamefont {Bates}}, \bibinfo {author} {\bibfnamefont
  {C.}~\bibnamefont {Stadler}}, \bibinfo {author} {\bibfnamefont
  {A.}~\bibnamefont {Hildebrand}},\ and\ \bibinfo {author} {\bibfnamefont
  {A.-K.}\ \bibnamefont {Classen}},\ }\bibfield  {title} {\bibinfo {title}
  {Calcium spikes, waves and oscillations in a large, patterned epithelial
  tissue},\ }\href@noop {} {\bibfield  {journal} {\bibinfo  {journal}
  {Scientific reports}\ }\textbf {\bibinfo {volume} {7}},\ \bibinfo {pages}
  {42786} (\bibinfo {year} {2017})}\BibitemShut {NoStop}%
\bibitem [{\citenamefont {He}\ \emph {et~al.}(2010)\citenamefont {He},
  \citenamefont {Wang}, \citenamefont {Tang},\ and\ \citenamefont
  {Montell}}]{he2010tissue}%
  \BibitemOpen
  \bibfield  {author} {\bibinfo {author} {\bibfnamefont {L.}~\bibnamefont
  {He}}, \bibinfo {author} {\bibfnamefont {X.}~\bibnamefont {Wang}}, \bibinfo
  {author} {\bibfnamefont {H.~L.}\ \bibnamefont {Tang}},\ and\ \bibinfo
  {author} {\bibfnamefont {D.~J.}\ \bibnamefont {Montell}},\ }\bibfield
  {title} {\bibinfo {title} {Tissue elongation requires oscillating
  contractions of a basal actomyosin network},\ }\href@noop {} {\bibfield
  {journal} {\bibinfo  {journal} {Nature cell biology}\ }\textbf {\bibinfo
  {volume} {12}},\ \bibinfo {pages} {1133} (\bibinfo {year}
  {2010})}\BibitemShut {NoStop}%
\bibitem [{\citenamefont {Martin}\ \emph {et~al.}(2009)\citenamefont {Martin},
  \citenamefont {Kaschube},\ and\ \citenamefont
  {Wieschaus}}]{martin2009pulsed}%
  \BibitemOpen
  \bibfield  {author} {\bibinfo {author} {\bibfnamefont {A.~C.}\ \bibnamefont
  {Martin}}, \bibinfo {author} {\bibfnamefont {M.}~\bibnamefont {Kaschube}},\
  and\ \bibinfo {author} {\bibfnamefont {E.~F.}\ \bibnamefont {Wieschaus}},\
  }\bibfield  {title} {\bibinfo {title} {Pulsed contractions of an
  actin--myosin network drive apical constriction},\ }\href@noop {} {\bibfield
  {journal} {\bibinfo  {journal} {Nature}\ }\textbf {\bibinfo {volume} {457}},\
  \bibinfo {pages} {495} (\bibinfo {year} {2009})}\BibitemShut {NoStop}%
\bibitem [{\citenamefont {P{\'a}lsson}\ \emph {et~al.}(1997)\citenamefont
  {P{\'a}lsson}, \citenamefont {Lee}, \citenamefont {Goldstein}, \citenamefont
  {Franke}, \citenamefont {Kessin},\ and\ \citenamefont
  {Cox}}]{palsson1997selection}%
  \BibitemOpen
  \bibfield  {author} {\bibinfo {author} {\bibfnamefont {E.}~\bibnamefont
  {P{\'a}lsson}}, \bibinfo {author} {\bibfnamefont {K.~J.}\ \bibnamefont
  {Lee}}, \bibinfo {author} {\bibfnamefont {R.~E.}\ \bibnamefont {Goldstein}},
  \bibinfo {author} {\bibfnamefont {J.}~\bibnamefont {Franke}}, \bibinfo
  {author} {\bibfnamefont {R.~H.}\ \bibnamefont {Kessin}},\ and\ \bibinfo
  {author} {\bibfnamefont {E.~C.}\ \bibnamefont {Cox}},\ }\bibfield  {title}
  {\bibinfo {title} {Selection for spiral waves in the social amoebae
  dictyostelium},\ }\href@noop {} {\bibfield  {journal} {\bibinfo  {journal}
  {Proceedings of the National Academy of Sciences}\ }\textbf {\bibinfo
  {volume} {94}},\ \bibinfo {pages} {13719} (\bibinfo {year}
  {1997})}\BibitemShut {NoStop}%
\bibitem [{\citenamefont {Turing}(1952)}]{turing1952}%
  \BibitemOpen
  \bibfield  {author} {\bibinfo {author} {\bibfnamefont {A.~M.}\ \bibnamefont
  {Turing}},\ }\bibfield  {title} {\bibinfo {title} {The chemical basis of
  morphogenesis},\ }\href@noop {} {\bibfield  {journal} {\bibinfo  {journal}
  {Philosophical Transactions of the Royal Society of London. Series B,
  Biological Sciences}\ }\textbf {\bibinfo {volume} {237}},\ \bibinfo {pages}
  {37} (\bibinfo {year} {1952})}\BibitemShut {NoStop}%
\bibitem [{\citenamefont {Gierer}\ and\ \citenamefont
  {Meinhardt}(1972)}]{gierer1972}%
  \BibitemOpen
  \bibfield  {author} {\bibinfo {author} {\bibfnamefont {A.}~\bibnamefont
  {Gierer}}\ and\ \bibinfo {author} {\bibfnamefont {H.}~\bibnamefont
  {Meinhardt}},\ }\bibfield  {title} {\bibinfo {title} {A theory of biological
  pattern formation},\ }\href@noop {} {\bibfield  {journal} {\bibinfo
  {journal} {Kybernetik}\ }\textbf {\bibinfo {volume} {12}},\ \bibinfo {pages}
  {30} (\bibinfo {year} {1972})}\BibitemShut {NoStop}%
\bibitem [{\citenamefont {Murray}(2003)}]{murray2003}%
  \BibitemOpen
  \bibfield  {author} {\bibinfo {author} {\bibfnamefont {J.~D.}\ \bibnamefont
  {Murray}},\ }\href@noop {} {\emph {\bibinfo {title} {Mathematical Biology:
  II: Spatial Models and Biomedical Applications}}},\ Vol.~\bibinfo {volume}
  {3}\ (\bibinfo  {publisher} {Springer},\ \bibinfo {year} {2003})\BibitemShut
  {NoStop}%
\bibitem [{\citenamefont {Hayden}\ \emph {et~al.}(2021)\citenamefont {Hayden},
  \citenamefont {Poss}, \citenamefont {De~Simone},\ and\ \citenamefont
  {Di~Talia}}]{hayden2021mathematical}%
  \BibitemOpen
  \bibfield  {author} {\bibinfo {author} {\bibfnamefont {L.~D.}\ \bibnamefont
  {Hayden}}, \bibinfo {author} {\bibfnamefont {K.~D.}\ \bibnamefont {Poss}},
  \bibinfo {author} {\bibfnamefont {A.}~\bibnamefont {De~Simone}},\ and\
  \bibinfo {author} {\bibfnamefont {S.}~\bibnamefont {Di~Talia}},\ }\bibfield
  {title} {\bibinfo {title} {Mathematical modeling of erk activity waves in
  regenerating zebrafish scales},\ }\href@noop {} {\bibfield  {journal}
  {\bibinfo  {journal} {Biophysical Journal}\ }\textbf {\bibinfo {volume}
  {120}},\ \bibinfo {pages} {4287} (\bibinfo {year} {2021})}\BibitemShut
  {NoStop}%
\bibitem [{\citenamefont {Kessler}\ and\ \citenamefont
  {Levine}(1993)}]{kessler1993pattern}%
  \BibitemOpen
  \bibfield  {author} {\bibinfo {author} {\bibfnamefont {D.~A.}\ \bibnamefont
  {Kessler}}\ and\ \bibinfo {author} {\bibfnamefont {H.}~\bibnamefont
  {Levine}},\ }\bibfield  {title} {\bibinfo {title} {Pattern formation in
  dictyostelium via the dynamics of cooperative biological entities},\
  }\href@noop {} {\bibfield  {journal} {\bibinfo  {journal} {Physical Review
  E}\ }\textbf {\bibinfo {volume} {48}},\ \bibinfo {pages} {4801} (\bibinfo
  {year} {1993})}\BibitemShut {NoStop}%
\bibitem [{\citenamefont {Levine}\ \emph {et~al.}(1996)\citenamefont {Levine},
  \citenamefont {Aranson}, \citenamefont {Tsimring},\ and\ \citenamefont
  {Truong}}]{levine1996positive}%
  \BibitemOpen
  \bibfield  {author} {\bibinfo {author} {\bibfnamefont {H.}~\bibnamefont
  {Levine}}, \bibinfo {author} {\bibfnamefont {I.}~\bibnamefont {Aranson}},
  \bibinfo {author} {\bibfnamefont {L.}~\bibnamefont {Tsimring}},\ and\
  \bibinfo {author} {\bibfnamefont {T.~V.}\ \bibnamefont {Truong}},\ }\bibfield
   {title} {\bibinfo {title} {Positive genetic feedback governs camp spiral
  wave formation in dictyostelium.},\ }\href@noop {} {\bibfield  {journal}
  {\bibinfo  {journal} {Proceedings of the National Academy of Sciences}\
  }\textbf {\bibinfo {volume} {93}},\ \bibinfo {pages} {6382} (\bibinfo {year}
  {1996})}\BibitemShut {NoStop}%
\bibitem [{\citenamefont {Grace}\ and\ \citenamefont
  {H{\"u}tt}(2015)}]{grace2015regulation}%
  \BibitemOpen
  \bibfield  {author} {\bibinfo {author} {\bibfnamefont {M.}~\bibnamefont
  {Grace}}\ and\ \bibinfo {author} {\bibfnamefont {M.-T.}\ \bibnamefont
  {H{\"u}tt}},\ }\bibfield  {title} {\bibinfo {title} {Regulation of
  spatiotemporal patterns by biological variability: General principles and
  applications to dictyostelium discoideum},\ }\href@noop {} {\bibfield
  {journal} {\bibinfo  {journal} {PLoS computational biology}\ }\textbf
  {\bibinfo {volume} {11}},\ \bibinfo {pages} {e1004367} (\bibinfo {year}
  {2015})}\BibitemShut {NoStop}%
\bibitem [{\citenamefont {Murray}\ \emph {et~al.}(1988)\citenamefont {Murray},
  \citenamefont {Maini},\ and\ \citenamefont {Tranquillo}}]{murray1988}%
  \BibitemOpen
  \bibfield  {author} {\bibinfo {author} {\bibfnamefont {J.~D.}\ \bibnamefont
  {Murray}}, \bibinfo {author} {\bibfnamefont {P.~K.}\ \bibnamefont {Maini}},\
  and\ \bibinfo {author} {\bibfnamefont {R.~T.}\ \bibnamefont {Tranquillo}},\
  }\bibfield  {title} {\bibinfo {title} {Mechanochemical models for generating
  biological pattern and form in development},\ }\href@noop {} {\bibfield
  {journal} {\bibinfo  {journal} {Physics Reports}\ }\textbf {\bibinfo {volume}
  {171}},\ \bibinfo {pages} {59} (\bibinfo {year} {1988})}\BibitemShut
  {NoStop}%
\bibitem [{\citenamefont {Bois}\ \emph {et~al.}(2011)\citenamefont {Bois},
  \citenamefont {J{\"u}licher},\ and\ \citenamefont {Grill}}]{bois2011pattern}%
  \BibitemOpen
  \bibfield  {author} {\bibinfo {author} {\bibfnamefont {J.~S.}\ \bibnamefont
  {Bois}}, \bibinfo {author} {\bibfnamefont {F.}~\bibnamefont {J{\"u}licher}},\
  and\ \bibinfo {author} {\bibfnamefont {S.~W.}\ \bibnamefont {Grill}},\
  }\bibfield  {title} {\bibinfo {title} {Pattern formation in active fluids},\
  }\href@noop {} {\bibfield  {journal} {\bibinfo  {journal} {Biophysical
  Journal}\ }\textbf {\bibinfo {volume} {100}},\ \bibinfo {pages} {445a}
  (\bibinfo {year} {2011})}\BibitemShut {NoStop}%
\bibitem [{\citenamefont {Banerjee}\ \emph {et~al.}(2015)\citenamefont
  {Banerjee}, \citenamefont {Utuje},\ and\ \citenamefont
  {Marchetti}}]{banerjee2015}%
  \BibitemOpen
  \bibfield  {author} {\bibinfo {author} {\bibfnamefont {S.}~\bibnamefont
  {Banerjee}}, \bibinfo {author} {\bibfnamefont {K.~J.}\ \bibnamefont
  {Utuje}},\ and\ \bibinfo {author} {\bibfnamefont {M.~C.}\ \bibnamefont
  {Marchetti}},\ }\bibfield  {title} {\bibinfo {title} {Propagating stress
  waves during epithelial expansion},\ }\href@noop {} {\bibfield  {journal}
  {\bibinfo  {journal} {Physical Review Letters}\ }\textbf {\bibinfo {volume}
  {114}},\ \bibinfo {pages} {228101} (\bibinfo {year} {2015})}\BibitemShut
  {NoStop}%
\bibitem [{\citenamefont {Boocock}\ \emph {et~al.}(2021)\citenamefont
  {Boocock}, \citenamefont {Hino}, \citenamefont {Ruzickova}, \citenamefont
  {Hirashima},\ and\ \citenamefont {Hannezo}}]{boocock2021theory}%
  \BibitemOpen
  \bibfield  {author} {\bibinfo {author} {\bibfnamefont {D.}~\bibnamefont
  {Boocock}}, \bibinfo {author} {\bibfnamefont {N.}~\bibnamefont {Hino}},
  \bibinfo {author} {\bibfnamefont {N.}~\bibnamefont {Ruzickova}}, \bibinfo
  {author} {\bibfnamefont {T.}~\bibnamefont {Hirashima}},\ and\ \bibinfo
  {author} {\bibfnamefont {E.}~\bibnamefont {Hannezo}},\ }\bibfield  {title}
  {\bibinfo {title} {Theory of mechanochemical patterning and optimal migration
  in cell monolayers},\ }\href@noop {} {\bibfield  {journal} {\bibinfo
  {journal} {Nature Physics}\ }\textbf {\bibinfo {volume} {17}},\ \bibinfo
  {pages} {267} (\bibinfo {year} {2021})}\BibitemShut {NoStop}%
\bibitem [{\citenamefont {Staddon}\ \emph {et~al.}(2022)\citenamefont
  {Staddon}, \citenamefont {Munro},\ and\ \citenamefont
  {Banerjee}}]{staddon2022pulsatile}%
  \BibitemOpen
  \bibfield  {author} {\bibinfo {author} {\bibfnamefont {M.~F.}\ \bibnamefont
  {Staddon}}, \bibinfo {author} {\bibfnamefont {E.~M.}\ \bibnamefont {Munro}},\
  and\ \bibinfo {author} {\bibfnamefont {S.}~\bibnamefont {Banerjee}},\
  }\bibfield  {title} {\bibinfo {title} {Pulsatile contractions and pattern
  formation in excitable actomyosin cortex},\ }\href@noop {} {\bibfield
  {journal} {\bibinfo  {journal} {PLoS Computational Biology}\ }\textbf
  {\bibinfo {volume} {18}},\ \bibinfo {pages} {e1009981} (\bibinfo {year}
  {2022})}\BibitemShut {NoStop}%
\bibitem [{\citenamefont {Armon}\ \emph {et~al.}(2018)\citenamefont {Armon},
  \citenamefont {Bull}, \citenamefont {Aranda-Diaz},\ and\ \citenamefont
  {Prakash}}]{armon2018ultrafast}%
  \BibitemOpen
  \bibfield  {author} {\bibinfo {author} {\bibfnamefont {S.}~\bibnamefont
  {Armon}}, \bibinfo {author} {\bibfnamefont {M.~S.}\ \bibnamefont {Bull}},
  \bibinfo {author} {\bibfnamefont {A.}~\bibnamefont {Aranda-Diaz}},\ and\
  \bibinfo {author} {\bibfnamefont {M.}~\bibnamefont {Prakash}},\ }\bibfield
  {title} {\bibinfo {title} {Ultrafast epithelial contractions provide insights
  into contraction speed limits and tissue integrity},\ }\href@noop {}
  {\bibfield  {journal} {\bibinfo  {journal} {Proceedings of the National
  Academy of Sciences}\ }\textbf {\bibinfo {volume} {115}},\ \bibinfo {pages}
  {E10333} (\bibinfo {year} {2018})}\BibitemShut {NoStop}%
\bibitem [{\citenamefont {Lin}\ \emph {et~al.}(2017)\citenamefont {Lin},
  \citenamefont {Li}, \citenamefont {Lan},\ and\ \citenamefont
  {Feng}}]{lin2017activation}%
  \BibitemOpen
  \bibfield  {author} {\bibinfo {author} {\bibfnamefont {S.-Z.}\ \bibnamefont
  {Lin}}, \bibinfo {author} {\bibfnamefont {B.}~\bibnamefont {Li}}, \bibinfo
  {author} {\bibfnamefont {G.}~\bibnamefont {Lan}},\ and\ \bibinfo {author}
  {\bibfnamefont {X.-Q.}\ \bibnamefont {Feng}},\ }\bibfield  {title} {\bibinfo
  {title} {Activation and synchronization of the oscillatory morphodynamics in
  multicellular monolayer},\ }\href@noop {} {\bibfield  {journal} {\bibinfo
  {journal} {Proceedings of the National Academy of Sciences}\ }\textbf
  {\bibinfo {volume} {114}},\ \bibinfo {pages} {8157} (\bibinfo {year}
  {2017})}\BibitemShut {NoStop}%
\bibitem [{\citenamefont {Serra-Picamal}\ \emph {et~al.}(2012)\citenamefont
  {Serra-Picamal}, \citenamefont {Conte}, \citenamefont {Vincent},
  \citenamefont {Anon}, \citenamefont {Tambe}, \citenamefont {Bazellieres},
  \citenamefont {Butler}, \citenamefont {Fredberg},\ and\ \citenamefont
  {Trepat}}]{serra2012mechanical}%
  \BibitemOpen
  \bibfield  {author} {\bibinfo {author} {\bibfnamefont {X.}~\bibnamefont
  {Serra-Picamal}}, \bibinfo {author} {\bibfnamefont {V.}~\bibnamefont
  {Conte}}, \bibinfo {author} {\bibfnamefont {R.}~\bibnamefont {Vincent}},
  \bibinfo {author} {\bibfnamefont {E.}~\bibnamefont {Anon}}, \bibinfo {author}
  {\bibfnamefont {D.~T.}\ \bibnamefont {Tambe}}, \bibinfo {author}
  {\bibfnamefont {E.}~\bibnamefont {Bazellieres}}, \bibinfo {author}
  {\bibfnamefont {J.~P.}\ \bibnamefont {Butler}}, \bibinfo {author}
  {\bibfnamefont {J.~J.}\ \bibnamefont {Fredberg}},\ and\ \bibinfo {author}
  {\bibfnamefont {X.}~\bibnamefont {Trepat}},\ }\bibfield  {title} {\bibinfo
  {title} {Mechanical waves during tissue expansion},\ }\href@noop {}
  {\bibfield  {journal} {\bibinfo  {journal} {Nature Physics}\ }\textbf
  {\bibinfo {volume} {8}},\ \bibinfo {pages} {628} (\bibinfo {year}
  {2012})}\BibitemShut {NoStop}%
\bibitem [{\citenamefont {Banerjee}\ and\ \citenamefont
  {Marchetti}(2019)}]{banerjee2019}%
  \BibitemOpen
  \bibfield  {author} {\bibinfo {author} {\bibfnamefont {S.}~\bibnamefont
  {Banerjee}}\ and\ \bibinfo {author} {\bibfnamefont {M.~C.}\ \bibnamefont
  {Marchetti}},\ }\bibfield  {title} {\bibinfo {title} {Continuum models of
  collective cell migration},\ }\href@noop {} {\bibfield  {journal} {\bibinfo
  {journal} {Cell Migrations: Causes and Functions}\ ,\ \bibinfo {pages} {45}}
  (\bibinfo {year} {2019})}\BibitemShut {NoStop}%
\bibitem [{\citenamefont {Odell}\ \emph {et~al.}(1981)\citenamefont {Odell},
  \citenamefont {Oster}, \citenamefont {Alberch},\ and\ \citenamefont
  {Burnside}}]{odell1981}%
  \BibitemOpen
  \bibfield  {author} {\bibinfo {author} {\bibfnamefont {G.~M.}\ \bibnamefont
  {Odell}}, \bibinfo {author} {\bibfnamefont {G.}~\bibnamefont {Oster}},
  \bibinfo {author} {\bibfnamefont {P.}~\bibnamefont {Alberch}},\ and\ \bibinfo
  {author} {\bibfnamefont {B.}~\bibnamefont {Burnside}},\ }\bibfield  {title}
  {\bibinfo {title} {The mechanical basis of morphogenesis: I. epithelial
  folding and invagination},\ }\href@noop {} {\bibfield  {journal} {\bibinfo
  {journal} {Developmental Biology}\ }\textbf {\bibinfo {volume} {85}},\
  \bibinfo {pages} {446} (\bibinfo {year} {1981})}\BibitemShut {NoStop}%
\bibitem [{\citenamefont {Heer}\ and\ \citenamefont
  {Martin}(2017)}]{heer2017tension}%
  \BibitemOpen
  \bibfield  {author} {\bibinfo {author} {\bibfnamefont {N.~C.}\ \bibnamefont
  {Heer}}\ and\ \bibinfo {author} {\bibfnamefont {A.~C.}\ \bibnamefont
  {Martin}},\ }\bibfield  {title} {\bibinfo {title} {Tension, contraction and
  tissue morphogenesis},\ }\href@noop {} {\bibfield  {journal} {\bibinfo
  {journal} {Development}\ }\textbf {\bibinfo {volume} {144}},\ \bibinfo
  {pages} {4249} (\bibinfo {year} {2017})}\BibitemShut {NoStop}%
\bibitem [{\citenamefont {Fernandez-Gonzalez}\ \emph
  {et~al.}(2009)\citenamefont {Fernandez-Gonzalez}, \citenamefont
  {de~Matos~Simoes}, \citenamefont {R{\"o}per}, \citenamefont {Eaton},\ and\
  \citenamefont {Zallen}}]{fernandez2009}%
  \BibitemOpen
  \bibfield  {author} {\bibinfo {author} {\bibfnamefont {R.}~\bibnamefont
  {Fernandez-Gonzalez}}, \bibinfo {author} {\bibfnamefont {S.}~\bibnamefont
  {de~Matos~Simoes}}, \bibinfo {author} {\bibfnamefont {J.-C.}\ \bibnamefont
  {R{\"o}per}}, \bibinfo {author} {\bibfnamefont {S.}~\bibnamefont {Eaton}},\
  and\ \bibinfo {author} {\bibfnamefont {J.~A.}\ \bibnamefont {Zallen}},\
  }\bibfield  {title} {\bibinfo {title} {Myosin ii dynamics are regulated by
  tension in intercalating cells},\ }\href@noop {} {\bibfield  {journal}
  {\bibinfo  {journal} {Developmental Cell}\ }\textbf {\bibinfo {volume}
  {17}},\ \bibinfo {pages} {736} (\bibinfo {year} {2009})}\BibitemShut
  {NoStop}%
\bibitem [{\citenamefont {Banerjee}\ \emph {et~al.}(2020)\citenamefont
  {Banerjee}, \citenamefont {Gardel},\ and\ \citenamefont
  {Schwarz}}]{banerjee2020}%
  \BibitemOpen
  \bibfield  {author} {\bibinfo {author} {\bibfnamefont {S.}~\bibnamefont
  {Banerjee}}, \bibinfo {author} {\bibfnamefont {M.~L.}\ \bibnamefont
  {Gardel}},\ and\ \bibinfo {author} {\bibfnamefont {U.~S.}\ \bibnamefont
  {Schwarz}},\ }\bibfield  {title} {\bibinfo {title} {The actin cytoskeleton as
  an active adaptive material},\ }\href@noop {} {\bibfield  {journal} {\bibinfo
   {journal} {Annual Review of Condensed Matter Physics}\ }\textbf {\bibinfo
  {volume} {11}},\ \bibinfo {pages} {421} (\bibinfo {year} {2020})}\BibitemShut
  {NoStop}%
\bibitem [{\citenamefont {Armon}\ \emph {et~al.}(2021)\citenamefont {Armon},
  \citenamefont {Bull}, \citenamefont {Moriel}, \citenamefont {Aharoni},\ and\
  \citenamefont {Prakash}}]{armon2021modeling}%
  \BibitemOpen
  \bibfield  {author} {\bibinfo {author} {\bibfnamefont {S.}~\bibnamefont
  {Armon}}, \bibinfo {author} {\bibfnamefont {M.~S.}\ \bibnamefont {Bull}},
  \bibinfo {author} {\bibfnamefont {A.}~\bibnamefont {Moriel}}, \bibinfo
  {author} {\bibfnamefont {H.}~\bibnamefont {Aharoni}},\ and\ \bibinfo {author}
  {\bibfnamefont {M.}~\bibnamefont {Prakash}},\ }\bibfield  {title} {\bibinfo
  {title} {Modeling epithelial tissues as active-elastic sheets reproduce
  contraction pulses and predict rip resistance},\ }\href@noop {} {\bibfield
  {journal} {\bibinfo  {journal} {Communications physics}\ }\textbf {\bibinfo
  {volume} {4}},\ \bibinfo {pages} {216} (\bibinfo {year} {2021})}\BibitemShut
  {NoStop}%
\bibitem [{\citenamefont {Boocock}\ \emph {et~al.}(2023)\citenamefont
  {Boocock}, \citenamefont {Hirashima},\ and\ \citenamefont
  {Hannezo}}]{boocock2023interplay}%
  \BibitemOpen
  \bibfield  {author} {\bibinfo {author} {\bibfnamefont {D.}~\bibnamefont
  {Boocock}}, \bibinfo {author} {\bibfnamefont {T.}~\bibnamefont {Hirashima}},\
  and\ \bibinfo {author} {\bibfnamefont {E.}~\bibnamefont {Hannezo}},\
  }\bibfield  {title} {\bibinfo {title} {Interplay between mechanochemical
  patterning and glassy dynamics in cellular monolayers},\ }\href
  {https://doi.org/10.1103/PRXLife.1.013001} {\bibfield  {journal} {\bibinfo
  {journal} {PRX Life}\ }\textbf {\bibinfo {volume} {1}},\ \bibinfo {pages}
  {013001} (\bibinfo {year} {2023})}\BibitemShut {NoStop}%
\bibitem [{\citenamefont {Noll}\ \emph {et~al.}(2017)\citenamefont {Noll},
  \citenamefont {Mani}, \citenamefont {Heemskerk}, \citenamefont {Streichan},\
  and\ \citenamefont {Shraiman}}]{noll2017}%
  \BibitemOpen
  \bibfield  {author} {\bibinfo {author} {\bibfnamefont {N.}~\bibnamefont
  {Noll}}, \bibinfo {author} {\bibfnamefont {M.}~\bibnamefont {Mani}}, \bibinfo
  {author} {\bibfnamefont {I.}~\bibnamefont {Heemskerk}}, \bibinfo {author}
  {\bibfnamefont {S.~J.}\ \bibnamefont {Streichan}},\ and\ \bibinfo {author}
  {\bibfnamefont {B.~I.}\ \bibnamefont {Shraiman}},\ }\bibfield  {title}
  {\bibinfo {title} {Active tension network model suggests an exotic mechanical
  state realized in epithelial tissues},\ }\href@noop {} {\bibfield  {journal}
  {\bibinfo  {journal} {Nature Physics}\ }\textbf {\bibinfo {volume} {13}},\
  \bibinfo {pages} {1221} (\bibinfo {year} {2017})}\BibitemShut {NoStop}%
\bibitem [{\citenamefont {Nagai}\ and\ \citenamefont
  {Honda}(2001)}]{nagai2001dynamic}%
  \BibitemOpen
  \bibfield  {author} {\bibinfo {author} {\bibfnamefont {T.}~\bibnamefont
  {Nagai}}\ and\ \bibinfo {author} {\bibfnamefont {H.}~\bibnamefont {Honda}},\
  }\bibfield  {title} {\bibinfo {title} {A dynamic cell model for the formation
  of epithelial tissues},\ }\href@noop {} {\bibfield  {journal} {\bibinfo
  {journal} {Philosophical Magazine B}\ }\textbf {\bibinfo {volume} {81}},\
  \bibinfo {pages} {699} (\bibinfo {year} {2001})}\BibitemShut {NoStop}%
\bibitem [{\citenamefont {Farhadifar}\ \emph {et~al.}(2007)\citenamefont
  {Farhadifar}, \citenamefont {R{\"o}per}, \citenamefont {Aigouy},
  \citenamefont {Eaton},\ and\ \citenamefont
  {J{\"u}licher}}]{farhadifar2007influence}%
  \BibitemOpen
  \bibfield  {author} {\bibinfo {author} {\bibfnamefont {R.}~\bibnamefont
  {Farhadifar}}, \bibinfo {author} {\bibfnamefont {J.-C.}\ \bibnamefont
  {R{\"o}per}}, \bibinfo {author} {\bibfnamefont {B.}~\bibnamefont {Aigouy}},
  \bibinfo {author} {\bibfnamefont {S.}~\bibnamefont {Eaton}},\ and\ \bibinfo
  {author} {\bibfnamefont {F.}~\bibnamefont {J{\"u}licher}},\ }\bibfield
  {title} {\bibinfo {title} {The influence of cell mechanics, cell-cell
  interactions, and proliferation on epithelial packing},\ }\href@noop {}
  {\bibfield  {journal} {\bibinfo  {journal} {Current Biology}\ }\textbf
  {\bibinfo {volume} {17}},\ \bibinfo {pages} {2095} (\bibinfo {year}
  {2007})}\BibitemShut {NoStop}%
\bibitem [{\citenamefont {Staple}\ \emph {et~al.}(2010)\citenamefont {Staple},
  \citenamefont {Farhadifar}, \citenamefont {R{\"o}per}, \citenamefont
  {Aigouy}, \citenamefont {Eaton},\ and\ \citenamefont
  {J{\"u}licher}}]{staple2010mechanics}%
  \BibitemOpen
  \bibfield  {author} {\bibinfo {author} {\bibfnamefont {D.~B.}\ \bibnamefont
  {Staple}}, \bibinfo {author} {\bibfnamefont {R.}~\bibnamefont {Farhadifar}},
  \bibinfo {author} {\bibfnamefont {J.-C.}\ \bibnamefont {R{\"o}per}}, \bibinfo
  {author} {\bibfnamefont {B.}~\bibnamefont {Aigouy}}, \bibinfo {author}
  {\bibfnamefont {S.}~\bibnamefont {Eaton}},\ and\ \bibinfo {author}
  {\bibfnamefont {F.}~\bibnamefont {J{\"u}licher}},\ }\bibfield  {title}
  {\bibinfo {title} {Mechanics and remodelling of cell packings in epithelia},\
  }\href@noop {} {\bibfield  {journal} {\bibinfo  {journal} {The European
  Physical Journal E}\ }\textbf {\bibinfo {volume} {33}},\ \bibinfo {pages}
  {117} (\bibinfo {year} {2010})}\BibitemShut {NoStop}%
\bibitem [{\citenamefont {Fletcher}\ \emph {et~al.}(2014)\citenamefont
  {Fletcher}, \citenamefont {Osterfield}, \citenamefont {Baker},\ and\
  \citenamefont {Shvartsman}}]{fletcher2014}%
  \BibitemOpen
  \bibfield  {author} {\bibinfo {author} {\bibfnamefont {A.~G.}\ \bibnamefont
  {Fletcher}}, \bibinfo {author} {\bibfnamefont {M.}~\bibnamefont
  {Osterfield}}, \bibinfo {author} {\bibfnamefont {R.~E.}\ \bibnamefont
  {Baker}},\ and\ \bibinfo {author} {\bibfnamefont {S.~Y.}\ \bibnamefont
  {Shvartsman}},\ }\bibfield  {title} {\bibinfo {title} {Vertex models of
  epithelial morphogenesis},\ }\href@noop {} {\bibfield  {journal} {\bibinfo
  {journal} {Biophysical journal}\ }\textbf {\bibinfo {volume} {106}},\
  \bibinfo {pages} {2291} (\bibinfo {year} {2014})}\BibitemShut {NoStop}%
\bibitem [{\citenamefont {Gustafson}\ \emph {et~al.}(2022)\citenamefont
  {Gustafson}, \citenamefont {Claussen}, \citenamefont {De~Renzis},\ and\
  \citenamefont {Streichan}}]{gustafson2022patterned}%
  \BibitemOpen
  \bibfield  {author} {\bibinfo {author} {\bibfnamefont {H.~J.}\ \bibnamefont
  {Gustafson}}, \bibinfo {author} {\bibfnamefont {N.}~\bibnamefont {Claussen}},
  \bibinfo {author} {\bibfnamefont {S.}~\bibnamefont {De~Renzis}},\ and\
  \bibinfo {author} {\bibfnamefont {S.~J.}\ \bibnamefont {Streichan}},\
  }\bibfield  {title} {\bibinfo {title} {Patterned mechanical feedback
  establishes a global myosin gradient},\ }\href@noop {} {\bibfield  {journal}
  {\bibinfo  {journal} {Nature Communications}\ }\textbf {\bibinfo {volume}
  {13}},\ \bibinfo {pages} {7050} (\bibinfo {year} {2022})}\BibitemShut
  {NoStop}%
\bibitem [{\citenamefont {Nyitrai}\ and\ \citenamefont
  {Geeves}(2004)}]{nyitrai2004adenosine}%
  \BibitemOpen
  \bibfield  {author} {\bibinfo {author} {\bibfnamefont {M.}~\bibnamefont
  {Nyitrai}}\ and\ \bibinfo {author} {\bibfnamefont {M.~A.}\ \bibnamefont
  {Geeves}},\ }\bibfield  {title} {\bibinfo {title} {Adenosine diphosphate and
  strain sensitivity in myosin motors},\ }\href@noop {} {\bibfield  {journal}
  {\bibinfo  {journal} {Philosophical Transactions of the Royal Society B:
  Biological Sciences}\ }\textbf {\bibinfo {volume} {359}},\ \bibinfo {pages}
  {1867} (\bibinfo {year} {2004})}\BibitemShut {NoStop}%
\bibitem [{\citenamefont {Uyeda}\ \emph {et~al.}(2011)\citenamefont {Uyeda},
  \citenamefont {Iwadate}, \citenamefont {Umeki}, \citenamefont {Nagasaki},\
  and\ \citenamefont {Yumura}}]{uyeda2011}%
  \BibitemOpen
  \bibfield  {author} {\bibinfo {author} {\bibfnamefont {T.~Q.}\ \bibnamefont
  {Uyeda}}, \bibinfo {author} {\bibfnamefont {Y.}~\bibnamefont {Iwadate}},
  \bibinfo {author} {\bibfnamefont {N.}~\bibnamefont {Umeki}}, \bibinfo
  {author} {\bibfnamefont {A.}~\bibnamefont {Nagasaki}},\ and\ \bibinfo
  {author} {\bibfnamefont {S.}~\bibnamefont {Yumura}},\ }\bibfield  {title}
  {\bibinfo {title} {Stretching actin filaments within cells enhances their
  affinity for the myosin ii motor domain},\ }\href@noop {} {\bibfield
  {journal} {\bibinfo  {journal} {PloS One}\ }\textbf {\bibinfo {volume} {6}},\
  \bibinfo {pages} {e26200} (\bibinfo {year} {2011})}\BibitemShut {NoStop}%
\bibitem [{\citenamefont {Burla}\ \emph {et~al.}(2019)\citenamefont {Burla},
  \citenamefont {Mulla}, \citenamefont {Vos}, \citenamefont
  {Aufderhorst-Roberts},\ and\ \citenamefont {Koenderink}}]{burla2019}%
  \BibitemOpen
  \bibfield  {author} {\bibinfo {author} {\bibfnamefont {F.}~\bibnamefont
  {Burla}}, \bibinfo {author} {\bibfnamefont {Y.}~\bibnamefont {Mulla}},
  \bibinfo {author} {\bibfnamefont {B.~E.}\ \bibnamefont {Vos}}, \bibinfo
  {author} {\bibfnamefont {A.}~\bibnamefont {Aufderhorst-Roberts}},\ and\
  \bibinfo {author} {\bibfnamefont {G.~H.}\ \bibnamefont {Koenderink}},\
  }\bibfield  {title} {\bibinfo {title} {From mechanical resilience to active
  material properties in biopolymer networks},\ }\href@noop {} {\bibfield
  {journal} {\bibinfo  {journal} {Nature Reviews Physics}\ }\textbf {\bibinfo
  {volume} {1}},\ \bibinfo {pages} {249} (\bibinfo {year} {2019})}\BibitemShut
  {NoStop}%
\bibitem [{\citenamefont {Staddon}\ \emph {et~al.}(2019)\citenamefont
  {Staddon}, \citenamefont {Cavanaugh}, \citenamefont {Munro}, \citenamefont
  {Gardel},\ and\ \citenamefont {Banerjee}}]{staddon2019mechanosensitive}%
  \BibitemOpen
  \bibfield  {author} {\bibinfo {author} {\bibfnamefont {M.~F.}\ \bibnamefont
  {Staddon}}, \bibinfo {author} {\bibfnamefont {K.~E.}\ \bibnamefont
  {Cavanaugh}}, \bibinfo {author} {\bibfnamefont {E.~M.}\ \bibnamefont
  {Munro}}, \bibinfo {author} {\bibfnamefont {M.~L.}\ \bibnamefont {Gardel}},\
  and\ \bibinfo {author} {\bibfnamefont {S.}~\bibnamefont {Banerjee}},\
  }\bibfield  {title} {\bibinfo {title} {Mechanosensitive junction remodeling
  promotes robust epithelial morphogenesis},\ }\href
  {https://doi.org/10.1016/j.bpj.2019.09.027} {\bibfield  {journal} {\bibinfo
  {journal} {Biophysical Journal}\ }\textbf {\bibinfo {volume} {117}},\
  \bibinfo {pages} {1739} (\bibinfo {year} {2019})}\BibitemShut {NoStop}%
\bibitem [{\citenamefont {Sawai}\ \emph {et~al.}(2005)\citenamefont {Sawai},
  \citenamefont {Thomason},\ and\ \citenamefont {Cox}}]{sawai2005}%
  \BibitemOpen
  \bibfield  {author} {\bibinfo {author} {\bibfnamefont {S.}~\bibnamefont
  {Sawai}}, \bibinfo {author} {\bibfnamefont {P.~A.}\ \bibnamefont
  {Thomason}},\ and\ \bibinfo {author} {\bibfnamefont {E.~C.}\ \bibnamefont
  {Cox}},\ }\bibfield  {title} {\bibinfo {title} {An autoregulatory circuit for
  long-range self-organization in dictyostelium cell populations},\ }\href@noop
  {} {\bibfield  {journal} {\bibinfo  {journal} {Nature}\ }\textbf {\bibinfo
  {volume} {433}},\ \bibinfo {pages} {323} (\bibinfo {year}
  {2005})}\BibitemShut {NoStop}%
\bibitem [{\citenamefont {Pertsov}\ \emph {et~al.}(1993)\citenamefont
  {Pertsov}, \citenamefont {Davidenko}, \citenamefont {Salomonsz},
  \citenamefont {Baxter},\ and\ \citenamefont {Jalife}}]{pertsov1993}%
  \BibitemOpen
  \bibfield  {author} {\bibinfo {author} {\bibfnamefont {A.~M.}\ \bibnamefont
  {Pertsov}}, \bibinfo {author} {\bibfnamefont {J.~M.}\ \bibnamefont
  {Davidenko}}, \bibinfo {author} {\bibfnamefont {R.}~\bibnamefont
  {Salomonsz}}, \bibinfo {author} {\bibfnamefont {W.~T.}\ \bibnamefont
  {Baxter}},\ and\ \bibinfo {author} {\bibfnamefont {J.}~\bibnamefont
  {Jalife}},\ }\bibfield  {title} {\bibinfo {title} {Spiral waves of excitation
  underlie reentrant activity in isolated cardiac muscle.},\ }\href@noop {}
  {\bibfield  {journal} {\bibinfo  {journal} {Circulation Research}\ }\textbf
  {\bibinfo {volume} {72}},\ \bibinfo {pages} {631} (\bibinfo {year}
  {1993})}\BibitemShut {NoStop}%
\bibitem [{\citenamefont {Lechleiter}\ \emph {et~al.}(1991)\citenamefont
  {Lechleiter}, \citenamefont {Girard}, \citenamefont {Peralta},\ and\
  \citenamefont {Clapham}}]{lechleiter1991}%
  \BibitemOpen
  \bibfield  {author} {\bibinfo {author} {\bibfnamefont {J.}~\bibnamefont
  {Lechleiter}}, \bibinfo {author} {\bibfnamefont {S.}~\bibnamefont {Girard}},
  \bibinfo {author} {\bibfnamefont {E.}~\bibnamefont {Peralta}},\ and\ \bibinfo
  {author} {\bibfnamefont {D.}~\bibnamefont {Clapham}},\ }\bibfield  {title}
  {\bibinfo {title} {Spiral calcium wave propagation and annihilation in
  xenopus laevis oocytes},\ }\href@noop {} {\bibfield  {journal} {\bibinfo
  {journal} {Science}\ }\textbf {\bibinfo {volume} {252}},\ \bibinfo {pages}
  {123} (\bibinfo {year} {1991})}\BibitemShut {NoStop}%
\bibitem [{\citenamefont {Reshodko}\ and\ \citenamefont
  {Bure{\v{s}}}(1975)}]{reshodko1975computer}%
  \BibitemOpen
  \bibfield  {author} {\bibinfo {author} {\bibfnamefont {L.}~\bibnamefont
  {Reshodko}}\ and\ \bibinfo {author} {\bibfnamefont {J.}~\bibnamefont
  {Bure{\v{s}}}},\ }\bibfield  {title} {\bibinfo {title} {Computer simulation
  of reverberating spreading depression in a network of cell automata},\
  }\href@noop {} {\bibfield  {journal} {\bibinfo  {journal} {Biological
  cybernetics}\ }\textbf {\bibinfo {volume} {18}},\ \bibinfo {pages} {181}
  (\bibinfo {year} {1975})}\BibitemShut {NoStop}%
\bibitem [{\citenamefont {Greenberg}\ and\ \citenamefont
  {Hastings}(1978)}]{greenberg1978spatial}%
  \BibitemOpen
  \bibfield  {author} {\bibinfo {author} {\bibfnamefont {J.~M.}\ \bibnamefont
  {Greenberg}}\ and\ \bibinfo {author} {\bibfnamefont {S.~P.}\ \bibnamefont
  {Hastings}},\ }\bibfield  {title} {\bibinfo {title} {Spatial patterns for
  discrete models of diffusion in excitable media},\ }\href@noop {} {\bibfield
  {journal} {\bibinfo  {journal} {SIAM Journal on Applied Mathematics}\
  }\textbf {\bibinfo {volume} {34}},\ \bibinfo {pages} {515} (\bibinfo {year}
  {1978})}\BibitemShut {NoStop}%
\bibitem [{\citenamefont {Gerhardt}\ \emph {et~al.}(1990)\citenamefont
  {Gerhardt}, \citenamefont {Schuster},\ and\ \citenamefont
  {Tyson}}]{gerhardt1990}%
  \BibitemOpen
  \bibfield  {author} {\bibinfo {author} {\bibfnamefont {M.}~\bibnamefont
  {Gerhardt}}, \bibinfo {author} {\bibfnamefont {H.}~\bibnamefont {Schuster}},\
  and\ \bibinfo {author} {\bibfnamefont {J.~J.}\ \bibnamefont {Tyson}},\
  }\bibfield  {title} {\bibinfo {title} {A cellular automaton model of
  excitable media: Ii. curvature, dispersion, rotating waves and meandering
  waves},\ }\href@noop {} {\bibfield  {journal} {\bibinfo  {journal} {Physica
  D: Nonlinear Phenomena}\ }\textbf {\bibinfo {volume} {46}},\ \bibinfo {pages}
  {392} (\bibinfo {year} {1990})}\BibitemShut {NoStop}%
\bibitem [{\citenamefont {Cavanaugh}\ \emph {et~al.}(2020)\citenamefont
  {Cavanaugh}, \citenamefont {Staddon}, \citenamefont {Munro}, \citenamefont
  {Banerjee},\ and\ \citenamefont {Gardel}}]{cavanaugh2020rhoa}%
  \BibitemOpen
  \bibfield  {author} {\bibinfo {author} {\bibfnamefont {K.~E.}\ \bibnamefont
  {Cavanaugh}}, \bibinfo {author} {\bibfnamefont {M.~F.}\ \bibnamefont
  {Staddon}}, \bibinfo {author} {\bibfnamefont {E.}~\bibnamefont {Munro}},
  \bibinfo {author} {\bibfnamefont {S.}~\bibnamefont {Banerjee}},\ and\
  \bibinfo {author} {\bibfnamefont {M.~L.}\ \bibnamefont {Gardel}},\ }\bibfield
   {title} {\bibinfo {title} {Rhoa mediates epithelial cell shape changes via
  mechanosensitive endocytosis},\ }\href@noop {} {\bibfield  {journal}
  {\bibinfo  {journal} {Developmental Cell}\ }\textbf {\bibinfo {volume}
  {52}},\ \bibinfo {pages} {152} (\bibinfo {year} {2020})}\BibitemShut
  {NoStop}%
\bibitem [{\citenamefont {Toettcher}\ \emph {et~al.}(2013)\citenamefont
  {Toettcher}, \citenamefont {Weiner},\ and\ \citenamefont
  {Lim}}]{toettcher2013}%
  \BibitemOpen
  \bibfield  {author} {\bibinfo {author} {\bibfnamefont {J.~E.}\ \bibnamefont
  {Toettcher}}, \bibinfo {author} {\bibfnamefont {O.~D.}\ \bibnamefont
  {Weiner}},\ and\ \bibinfo {author} {\bibfnamefont {W.~A.}\ \bibnamefont
  {Lim}},\ }\bibfield  {title} {\bibinfo {title} {Using optogenetics to
  interrogate the dynamic control of signal transmission by the ras/erk
  module},\ }\href@noop {} {\bibfield  {journal} {\bibinfo  {journal} {Cell}\
  }\textbf {\bibinfo {volume} {155}},\ \bibinfo {pages} {1422} (\bibinfo {year}
  {2013})}\BibitemShut {NoStop}%
\bibitem [{\citenamefont {P{\'e}rez-Verdugo}\ and\ \citenamefont
  {Banerjee}(2022)}]{perez2022tension}%
  \BibitemOpen
  \bibfield  {author} {\bibinfo {author} {\bibfnamefont {F.}~\bibnamefont
  {P{\'e}rez-Verdugo}}\ and\ \bibinfo {author} {\bibfnamefont {S.}~\bibnamefont
  {Banerjee}},\ }\bibfield  {title} {\bibinfo {title} {Tension remodeling
  controls topological transitions and fluidity in epithelial tissues},\
  }\href@noop {} {\bibfield  {journal} {\bibinfo  {journal} {arXiv preprint
  arXiv:2211.05591}\ } (\bibinfo {year} {2022})}\BibitemShut {NoStop}%
\end{thebibliography}
\end{document}